\documentclass[equations,11pt]{article}
\usepackage{amssymb,amsmath,amsthm}
\usepackage{eucal}
\usepackage[dvips]{epsfig}

\renewcommand{\theequation}{\arabic{equation}}
\begin{document}
\bibliographystyle{plain}
\def\m@th{\mathsurround=0pt}
\mathchardef\bracell="0365 
\def\upbrall{$\m@th\bracell$}
\def\undertilde#1{\mathop{\vtop{\ialign{##\crcr
    $\hfil\displaystyle{#1}\hfil$\crcr
     \noalign
     {\kern1.5pt\nointerlineskip}
     \upbrall\crcr\noalign{\kern1pt
   }}}}\limits}
\def\theequation{\arabic{section}.\arabic{equation}}
\newcommand{\pp}{\partial}
\newcommand{\ar}{\alpha}
\newcommand{\aar}{\bar{a}}
\newcommand{\bb}{\beta}
\newcommand{\gm}{\gamma}
\newcommand{\Gm}{\Gamma}
\newcommand{\en}{\epsilon}
\newcommand{\ven}{\varepsilon}
\newcommand{\dd}{\delta}
\newcommand{\sg}{\sigma}
\newcommand{\kp}{\kappa}
\newcommand{\ld}{\lambda}
\newcommand{\vf}{\varphi}
\newcommand{\Ups}{\Upsilon}
\newcommand{\oa}{\omega}
\newcommand{\hf}{\frac{1}{2}}
\newcommand{\be}{\begin{equation}}
\newcommand{\ee}{\end{equation}}
\newcommand{\bea}{\begin{eqnarray}}
\newcommand{\eea}{\end{eqnarray}}
\newcommand{\bse}{\begin{subequations}}
\newcommand{\beast}{\begin{eqnarray*}}
\newcommand{\east}{\end{eqnarray*}}
\newcommand{\ese}{\end{subequations}}
\newcommand{\nn}{\nonumber}
\newcommand{\bR}{\bar{R}}
\newcommand{\bP}{\bar{\Phi}}
\newcommand{\bS}{\bar{S}}
\newcommand{\mbe}{{\boldsymbol e}}
\newcommand{\bp}{{\boldsymbol p}}
\newcommand{\bn}{{\boldsymbol n}}
\newcommand{\boa}{{\boldsymbol \omega}}
\newcommand{\bet}{{\boldsymbol \eta}}
\newcommand{\bW}{\bar{W}}
\newcommand{\sn}{{\rm sn}}
\newcommand{\wh}{\widehat}
\newcommand{\ol}{\overline}
\newcommand{\wt}{\widetilde}
\newcommand{\ut}{\undertilde}
 \newcommand{\bU}{\bf U}
 \newcommand{\pl}{\partial}
 \newcommand{\ddp}{\frac{\partial}{\partial p}}
 \newcommand{\ddq}{\frac{\partial}{\partial q}}
 \newcommand{\ddr}{\frac{\partial}{\partial r}}
 \newcommand{\Ld}{{\bf \Lambda}}
 \newcommand{\tLd}{\,^{t\!}{\bf \Lambda}}
 \newcommand{\I}{{\bf I}}
 \newcommand{\tII}{\,^{t\!}{\bf I}}
 \newcommand{\tuk}{\,^{t\!}{\bf u}_k}
 \newcommand{\tul}{\,^{t\!}{\bf u}_\ell}
 \newcommand{\tcl}{\,^{t\!}{\bf c}_{\ell}}
 \newcommand{\ssk}{\sigma_{k^\prime}}
 \newcommand{\ssl}{\sigma_{\ell^\prime}}
 \newcommand{\ddint}{\int_\Gamma d\ld(\ell) }
 \def\hypotilde#1#2{\vrule depth #1 pt width 0pt{\smash{{\mathop{#2}
 \limits_{\displaystyle\widetilde{}}}}}}
 \def\hypohat#1#2{\vrule depth #1 pt width 0pt{\smash{{\mathop{#2}
 \limits_{\displaystyle\widehat{}}}}}}
 \def\hypo#1#2{\vrule depth #1 pt width 0pt{\smash{{\mathop{#2}
 \limits_{\displaystyle{}}}}}}
 
\newtheorem{theorem}{Theorem}[section]
\newtheorem{lemma}{Lemma}[section]
\newtheorem{cor}{Corollary}[section]
\newtheorem{prop}{Proposition}[section]
\newtheorem{definition}{Definition}[section]
\newtheorem{conj}{Conjecture}[section]
 
 \begin{center}
 {\large{\bf The Discrete and Continuous Painlev\'e VI Hierarchy\\ 
 and the Garnier Systems}} 
 \vspace{1.2cm}
 
 F.W. Nijhoff ~and~ A.J. Walker\vspace{.15cm} \\
 {\it Department of Applied Mathematics \\
 The University of Leeds, Leeds LS2 9JT, UK}\\
 \end{center}
 \vspace{1.4cm}
 
 \centerline{\bf Abstract}
 \vspace{.2cm}

We present a general scheme to derive  higher-order members of 
the Painlev\'e VI (PVI) hierarchy of ODE's as well as their difference 
analogues. The derivation is based on a discrete structure that sits 
on the background of the PVI equation and that consists of a system 
of partial difference equations on a multidimensional lattice. 
The connection with the isomonodromic Garnier systems 
is discussed. 
\vspace{.5cm}

\vfill 
\noindent \underline{Key Words}: Discrete Painlev\'e Equations;
Painlev\'e Transcendents; Garnier systems; Similarity Reduction; 
Ordinary \& Partial Difference Equations; Isomonodromic Deformation 
Problems.
 
\setcounter{page}{0}
\pagebreak

\section{Introduction}
\setcounter{equation}{0}

In recent years there has been a growing interest in discrete 
analogues of the famous Painlev\'e equations, i.e. nonlinear 
nonautonomous ordinary difference equations tending to the continuous 
Painlev\'e equations in a well-defined limit and which are 
integrable in their own right, cf. \cite{Carg}. Even though 
the qualitative features of the solutions of these systems are not 
yet fully understood, nonetheless in most of the known examples the 
main ingredients of their integrability have been exhibited. 
Recently, a classification of continuous as well as discrete 
Painlev\'e equations in terms of the root systems associated with 
affine Weyl groups, has been proposed on the basis of the 
singularities of the rational surfaces of their initial conditions 
and their blowings-up, cf. \cite{Sakai}. 

In a recent paper, \cite{NRGO}, we established a connection between 
the continuous Painlev\'e VI (P$_{\rm VI}$) equation and a 
non-autonomous ordinary difference equation depending on four arbitrary 
parameters. This novel example of a discrete Painlev\'e equation 
arises on the one hand as the nonlinear addition formula for the 
P$_{\rm VI}$ transcendents, in fact what is effectively a superposition 
formula for its B\"acklund-Schlesinger transforms, on the other 
hand from the similarity reduction on the lattice (cf. \cite{NP,DIGP}), 
of a system of partial difference equations associated with the 
lattice KdV family.  
In subsequent papers, \cite{NJH,NHJ}, some more results
on these systems were established, namely the existence of the Miura 
chain and the discovery of a novel Schwarzian PDE generating the 
entire (Schwarzian) KdV hierarchy of nonlinear evolution equations and 
whose similarity reduction is exactly the 
P$_{\rm VI}$ equation, this being the first example of an integrable 
scalar PDE that reduces to full P$_{\rm VI}$ with arbitrary parameters. 

In the present note we extend these results to multi-dimensional 
systems associated with higher-order generalisations of the 
P$_{\rm VI}$ equation. Already in \cite{NRGO} we noted that 
the similarity reduction of the lattice KdV system could be 
generalised in a natural way to higher-order differential 
and difference equations, 
without, however, clarifying in detail the nature of such equations. 
What we will argue here is that, in fact, such equations constitute 
what one could call the Painlev\'e VI {\it hierarchy} and its 
discrete counterpart. Whilst the idea of constructing hierarchies of 
Painlev\'e equations by exploiting the similarity reductions of 
hierarchies of nonlinear evolution equations of KdV type is at least 
two decades old, cf. \cite{FN}, the issue has gained renewed interest 
in recent years, cf. e.g. \cite{Kudry}--\cite{Pick}, because of the 
hypothetical possibility that these hierarchies of higher-order 
Painlev\'e equations yield new transcendents. 
Evidence to that effect might be given by the asymptotic analysis of 
the higher-order equations, since they seem to be governed by 
hyper-elliptic functions rather than elliptic ones as is the case for 
the original Painlev\'e equations, \cite{Kitaev}. 

Most of the existing results on hierarchies of discrete and continuous 
Painlev\'e equations are restricted to the examples of 
P$_{\rm I}$ and P$_{\rm II}$ hierarchies, since only in these cases 
it is clear what hierarchies of nonlinear evolution  equations 
should be taken as the starting point for their construction. In the 
case of the other Painlev\'e equations, notably P$_{\rm VI}$, it has 
been less clear what to take as a starting point for the construction 
of its hierarchy. 
With the results of \cite{NRGO,NJH,NHJ} we are now well-equipped to 
tackle this problem, and in the present paper we outline the basic 
construction of the equations in the discrete as well as continuous 
P$_{\rm VI}$ hierarchy. 
In fact, we shall demonstrate that the lattice KdV system can 
be naturally embedded in a multidimensional lattice system achieving 
the higher-order reductions by including more terms in the relevant 
similarity constraint which provokes the coupling between the 
various lattice directions. 

It should be noted that in a sense higher-order P$_{\rm VI}$ systems 
already were constructed by R. Garnier in his celebrated paper of 
1912, \cite{Garnier}, extending the original approach of R. Fuchs 
who was the first in \cite{Fuchs} to find P$_{\rm VI}$ arising from 
the isomonodromic deformation of a second-order linear differential 
equation. We will conclude our paper with a discussion of these 
Garnier systems, which in view of the recent interest in algebraic 
solutions of P$_{\rm VI}$, cf. e.g. \cite{Hitch}-\cite{Dubr}, 
deserve in our opinion some renewed attention.

\section{The Discrete PVI Hierarchy}
\setcounter{equation}{0}
 
In \cite{NRGO}, following earlier work e.g. \cite{NP,DIGP}, cf. also 
\cite{Carg}, a coherent framework was developed in which the similarity 
reduction of both discrete as well as continuous equations associated 
with the lattice KdV family were treated. Surprisingly, from these 
reductions the full P$_{\rm VI}$ equation for 
arbitrary parameters emerged together with a four-parameter discrete 
equation, i.e. a discrete Painlev\'e equation. 
{}From the treatment  of \cite{NRGO} it was evident how to 
to extend the lattice system of partial difference equations and their 
similarity constraints leading to the reductions to P$_{\rm VI}$. 
Here we describe explicitely this higher-dimensional lattice system 
and discuss their explicit reductions. 

The lattice KdV family of equations contains many related equations 
such as the lattice Schwarzian KdV, the lattice modified KdV (mKdV) 
and the actual lattice KdV equations. We concentrate here on one 
member of this family only, namely the lattice mKdV equation:
\begin{equation}\label{eq:dMKdV} 
p\,v_{n,m}v_{n,m+1}+q\,v_{n,m+1}v_{n+1,m+1}=
q\,v_{n,m}v_{n+1,m}+p\,v_{n+1,m}v_{n+1,m+1} 
\end{equation} 
cf. e.g. \cite{KDV}, with discrete independent variables $n,m$ and 
depending on additional 
parameters of the equation $p,q$, i.e. the {\it lattice parameters}. 
As was pointed out earlier, cf. \cite{KDV}, the lattice equation 
(\ref{eq:dMKdV}) actually represents a {\it compatible parameter-family 
of partial difference equations}: namely, we can embed the equation 
(\ref{eq:dMKdV}) into a multidimensional lattice by imposing a copy 
of (\ref{eq:dMKdV}) with different parameters on any two-dimensional 
sublattice, identifying each lattice direction with a corresponding 
lattice parameter $p_i\in\mathbb C$ in which direction 
the sites are labelled by discrete variables $n_i$ (noting that these 
are not necessarily integers, but {\it shift} by units, i.e. 
$n_i\in\theta_i+\mathbb Z$, $\theta_i\in\mathbb C$).  
Thus, combining two different lattice directions, labelled by $(i,j)$ 
we can write the lattice equation (\ref{eq:dMKdV}) on the 
corresponding sublattice as 
\begin{equation}\label{eq:genMKdV} 
p_i vv^j+p_j v^j v^{ij}=
p_j vv^i+p_i v^i v^{ij}
\end{equation} 
in which we use the right superscripts $i,j$ to denote the shifts in 
the corresponding directions, whereas we will use left subscripts $i,j$ 
denote shifts in the reverse direction, i.e. 
\[ v=v(\bn;\bp)\      \ ,\     \ 
v^j=T_j v(\bn;\bp)=v(\bn+\mbe_j;
\bp)\     \ , \       \  
\,_{j\!}v=T_j^{-1}v(\bn;\bp)=
v(\bn-\mbe_j;\bp)\   , \]
where $\bn$ denotes the vector of the discrete variables $n_i$, for all 
lattice directsion labelled by $i$, each corresponding to the component 
$p_i$ of the vector $\bp$ of lattice parameters. We use the 
vector $\mbe_j$ to denote the vector with single nonzero entry 
equal to unity in its $j^{\rm th}$ component. 

The consistency of the lattice equation (\ref{eq:genMKdV}) along the 
multi-dimensional lattice follows from the diagram of Figure 1: 
considering the three-dimensional sublattice with elementary directions 
$\{ \mbe_1,\mbe_2,\mbe_3 \}$ then on each elementary 
cube in this lattice the iteration of initial data proceeds along the 
six faces of this cube, on each of which we have an equation of the 
form (\ref{eq:genMKdV}). Thus, starting from initial data $v$, $v^1$, 
$v^2$, $v^3$ we can then uniquely calculate the values of $v^{12}$, 
$v^{13}$ and $v^{23}$ by using the equation. However, proceeding 
further there are in principle three different ways to calculate the 
value of $v^{123}$, unless the equation satisfies (as is the case for 
the equation (\ref{eq:genMKdV})) the special property that these 
three different ways of calculating this point actually lead to one 
and the same value. It is indeed at this point that the consistency of 
the embedding of the lattice MKdV into the multidimensional lattice 
is tested. In fact, by a straightforward calculation we find 
that this value is given by 
\[  v^{ijk} = \frac{(p_i^2-p_k^2)p_jv^iv^k+(p_j^2-p_i^2)p_kv^jv^i+
(p_k^2-p_j^2)p_iv^jv^k}{(p_i^2-p_k^2)p_jv^j+(p_j^2-p_i^2)p_kv^k 
+(p_k^2-p_j^2)p_iv^i}\     \ ,\     \ i,j,k=1,2,3   \] 
(which is clearly invariant for any permutation of the labels $ijk$), 
independent of the way in which we calculate this value! Thus, the 
equation (\ref{eq:genMKdV}) can be simultaneously imposed on functions 
$v(n_1,n_2,n_3,\dots)$ of the lattice sites. This is precisely the 
discrete analogue of the hierarchy of commuting higher-order flows of 
the (modified) KdV equation! 
\begin{figure}[h]
\centering
\epsfig{file=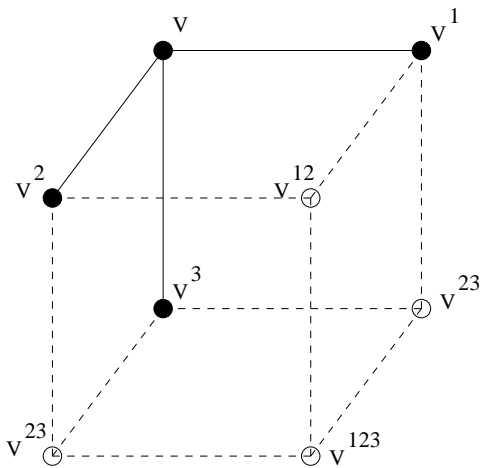,width=2.5in}
\caption{\it{Consistency of the lattice equation}.}
\end{figure}
\\
As a consequence of this compatibility we will call the system 
(\ref{eq:genMKdV}) a {\it holonomic system of partial difference 
equations}. 

Now, we turn to the issue of the symmetry reduction of the 
multidimensional lattice in the sense of \cite{DIGP}. 
It follows from the general framework of \cite{NRGO} that the 
similarity constraint for the multidimensional lattice 
MKdV system is as follows:
\begin{equation}\label{eq:pviconstr}
\sum_i\,n_i a_i = \mu-\nu\      \ ,\      \ 
\nu=\ld (-1)^{\sum_i n_i} \     , 
\end{equation} 
$\mu$ and $\ld$ being constants, and in which the variables $a_i$ 
are given by 
\begin{equation}\label{eq:a}  
a_i\equiv\frac{v^i-\,_{i\!}v}{v^i+\,_{i\!}v}\   . 
\end{equation} 
The sum in (\ref{eq:pviconstr}) is over all the $i$ labelling the 
lattice directions, the choice of which decides the order 
of the reduction. To analyse the reduction we need a 
number of relations for the objects $a_i$ which follow from 
(\ref{eq:genMKdV}), namely 
\bea\label{eq:arel}
1+a_j^i&=& \frac{(p_i X_{ij}-p_j)(a_j+1)+2p_j}
{p_i x_{ij}+p_j} \      \ ,\     \ i\neq j  \label{eq:arela} \\
a_i &=& \frac{p_j\:_iX_{ij}\:X_{ij}+p_i(\,_iX_{ij}-X_{ij})-p_j}
{p_j\:_iX_{ij}\:X_{ij}-p_i(\,_iX_{ij}+X_{ij})+p_j}  \nonumber \\ 
&=& \frac{-p_j\:_ix_{ij}\:x_{ij}-p_i(\,_ix_{ij}-x_{ij})+p_j}
{p_j\:_ix_{ij}\:x_{ij}+p_i(\,_ix_{ij}+x_{ij})+p_j}\       \ ,\      \ 
i\neq j  \label{eq:arelb} 
\eea 
in terms of the following variables:
\bea\label{eq:xydef}
x_{ij}\equiv\frac{v}{v^{ij}}\      \ &,&\      \ 
\,_ix_{ij}\equiv T_i^{-1}x_{ij}=\frac{\,_iv}{v^j}\ \ , \\ 
X_{ij}\equiv\frac{v^i}{v^j}\       \ &,&\      \ 
\,_iX_{ij}\equiv T_i^{-1}X_{ij}=\frac{v}{\,_iv^j}   . 
\eea 
The variables $x_{ij}=x_{ji}$ and $X_{ij}=1/X_{ji}$ are not 
independent, but related via:
\begin{equation}\label{eq:xyrel}
X_{ij}=\frac{p_i x_{ij}+p_j}{p_j x_{ij}+p_i}
\       \  \Leftrightarrow\       \   
x_{ij}=\frac{-p_i X_{ij}+p_j}{p_j X_{ij}-p_i}\  , 
\end{equation}
as well as
\begin{equation}\label{eq:axy}
\frac{T_i^{-1}x_{ij}}{X_{ij}}=
\frac{1-a_i}{1+a_i}\    .  
\end{equation} 
We note that since the left-hand side of (\ref{eq:arelb}) depends only 
on the label $i$ but not on $j$, for fixed $i$ this represents a set 
of $N-2$ coupled first-order ordinary difference equations with 
respect to the shift in the discrete variable $n_i$ between 
the $N-1$ variables $X_{ij}$, $j\neq i$. Furthermore, the relations 
(\ref{eq:arela}), for the same fixed label $i$, provide us with a set 
of $N-1$ first-order relations between the variables $a_j$, $j\neq i$, 
and thus together with the similarity constraint (\ref{eq:pviconstr}) 
where $a_i$ is substituted by (\ref{eq:arelb}) we obtain a set of 
$2(N-1)$ first-order nonlinear ordinary difference equations for the 
$2(N-1)$ variables $X_{ij}$, $a_j$, $j\neq i$, which together form our 
higher-order discrete system. In the next section we will explicitely 
disentangle this coupled system in the cases $N=2$ and $N=3$.

The continuous equation for the PVI hierarchy derive from the 
differential equations with respect to the lattice parameters 
$p_i$, which read: 
\begin{equation}\label{eq:dpv}
-p_i\frac{\pl}{\pl p_i} \log v=n_i a_i\    .
\end{equation} 
It can be shown that the differential relations (\ref{eq:dpv}) are 
actually compatible not only amongst themselves, but also with the 
the discrete equations on the lattice (\ref{eq:genMKdV}), i.e. the 
discrete and continuous flows are commuting: 
\[ \frac{\pl}{\pl p_i}\left(\frac{\pl v}{\pl p_j}\right)= 
\frac{\pl}{\pl p_j}\left(\frac{\pl v}{\pl p_i}\right)\        \ 
,\        \ \frac{\pl v^i}{\pl p_j}=T_i\left( \frac{\pl v}{\pl p_j}
\right)\   . \] 
This can actually be demonstrated by explicit calculation exploiting  
the discrete relations (\ref{eq:arel}), but we will not give the 
details here (which follow closely the pattern of calculations of 
\cite{NRGO}). Thus, we have a coherent framework of a large 
multidimensional system of equations with discrete (in terms of the 
variables $n_i$) as well as continuous (in terms of the parameters 
$p_i$) commuting flows, in terms of which compatible equations of 
three different types (partial difference, differential-difference 
and partial differential) figure in one and the same framework: the 
partial difference equations are precisely the lattice equations 
(\ref{eq:genMKdV}), the differential-difference equations are the 
relations (\ref{eq:dpv}), whilst for the partial differential equations 
in the scheme we refer to our recent paper \cite{NHJ}. Here we will 
focus now on the reductions under the symmetry constraint 
(\ref{eq:pviconstr}) in order to derive closed-from ODE's in terms of 
the lattice parameter $p_i$. To make this reduction explicit we 
use (\ref{eq:dpv}) in combination with (\ref{eq:arel})-(\ref{eq:axy}) 
to obtain differential relations for the $a_i$, namely 
\begin{equation}\label{eq:pla}
\frac{\pl a_j}{\pl p_i} = \frac{n_i p_j}{p_j^2-p_i^2}
\left[ (1+a_i)(1-a_j)X_{ji}-(1+a_j)(1-a_i)X_{ij} 
\right] \   ,
\end{equation}
as well as the following relations for the reduced 
variables $X_{ij}$ 
\bea \label{eq:plX} 
\mu+\nu+p_i\frac{\pl}{\pl p_i} \log\,X_{ij}&=&n_i\mathcal{X}_{ji}a_i 
+\sum_{k\neq i} n_k\mathcal{X}_{ik} a_k   \\
&& + n_i\frac{p_ip_j}{p_i^2-p_j^2}(X_{ji}-X_{ij})  
+ \sum_{k\neq i} n_k\frac{p_kp_i}{p_k^2-p_i^2}(X_{ik}-X_{ki})  \nn
\eea
in which we have abbreviated 
\begin{equation}
\mathcal{X}_{ij}\equiv 
\frac{(p_iX_{ij}-p_j)(p_j-p_iX_{ji})}{p_j^2-p_i^2} = -\mathcal{X}_{ji}
\   . 
\end{equation} 
Using (\ref{eq:pla}) in conjunction with (\ref{eq:plX}) 
and using the similarity constraint (\ref{eq:pviconstr}) to eliminate 
the $a_i$, we obtain a coupled first-order system of differential  
equations w.r.t. the independent variable $t_i=p_i^2$ in terms of 
the $2N-2$ variables ~$a_k$, $X_{ik}$~, ($k\neq i$). Solving the 
variables $a_k$ from the linear system given by the equations 
(\ref{eq:plX}) and inserting them into (\ref{eq:pla}) we obtain a 
coupled set of second-order nonlinear differential equations for the 
variables $X_{ik}$.

\section{Special Cases: N=2, N=3}
\setcounter{equation}{0}

We will now analyse the basic relations of the general framework 
presented in the previous section in the cases $N=2,3$ only in order 
to arrive at slightly more explicit equations, demonstrating that the
reduction leads to ordinary difference equations (in the discrete case) 
or to ordinary differential equations (in the continuous case). 

\subsection*{N=2:}

We will be very brief about the two-dimensional case $N=2$ which was 
the main subject of study in the earlier paper \cite{NRGO}. There 
the compatibility of the similarity constraint and the lattice 
equation was stated, and the various relations resulting from 
(\ref{eq:arel}) were already written down. Using in this case the 
slightly simpler notation: 
\[ a_1=a\   \ ,\   \ a_2=b\   \ ,\   \ 
x_{12}=x\    \ ,\   \ X_{12}=X\    \] 
and using the discrete independent variables $n_1=n$, $n_2=m$, as 
well as the lattice parameters $p_1=p$, $p_2=q$, we derived the 
second order nonlinear non-autonomous difference equation: 
\bea\label{eq:GDP}
&& \frac{2(n+1)}{1-y_{n+1}y_{n}} +
\frac{2n}{1-y_{n}y_{n-1}} = 
\mu+\ld(-1)^n+ 2n+1 + \nn \\
&+& 
\frac{ (\mu-\ld(-1)^n)(r^2-1)y_n + r(1-y_n^2)
\left[ (n+\frac{1}{2}) - (m+\frac{1}{2}) 
(-1)^n\right] } 
{ (r+y_{n})(1+ry_{n})}\   ,
\eea 
(using a slightly different notation from the one of \cite{NRGO}), 
where $r=p/q$ and where the variables $y_n$ are related to the 
$X$ and $x$ by the prescription: $y_{2n}=x(2n)$ for the even sites, 
whilst $y_{2n+1}=-1/X(2n+1)$ for the odd lattice sites (the latter 
choice being mainly motivated by the wish to cast the equation into 
a convenient shape). It was pointed out in \cite{NRGO} that whilst 
a continuum limit of (\ref{eq:GDP}) yields the P$_{\rm V}$ equation, 
its general solution can be expressed in terms of P$_{\rm VI}$ 
transcendents (noting its dependence on 
four arbitrary parameters, $\mu$, $\ld$, $r$ and $m$). 

The continuous equation for the variable $X$ in terms of the 
lattice parameter $p$ as independent variable in this case reads: 
\bea\label{eq:altPVI}
&&p(p^2-q^2)^2X(qX-p)(pX-q)\frac{\pl ^2X}{\pl p^2}=\nn \\ 
&&=\frac{1}{2}p(p^2-q^2)^2\left[ pq(3X^2+1)-2(p^2+q^2)X\right]
\left(\frac{\pl X}{\pl p}\right)^2 + \nn\\ 
&&~+(q^2-p^2)\left[ 2p^2X(pX-q)(qX-p)+(q^2-p^2)^2X^2\right] 
\frac{\pl X}{\pl p}  \nn \\
&&~+\frac{1}{2}q\left[ (\ar X^2-\bb)(pX-q)^2(qX-p)^2 
+(p^2-q^2)X^2\left((\gm-1)(qX-p)^2-(\dd-1)(pX-q)^2\right)\right]\  ,  
\nn \\ \eea 
and it is not difficult to show that this is actually the 
P$_{\rm VI}$ equation through the identification $w(t)=pX(p)$, 
where $t=p^2$, and setting $q=1$, leading to 
\bse\label{eq:pvi}\bea\label{eq:PVI}
\frac{d^2w}{dt^2}=&& \frac{1}{2}\left( \frac{1}{w}+\frac{1}{w-1}+
\frac{1}{w-t}\right)\left(\frac{dw}{dt}\right)^2 -
\left(\frac{1}{t}+\frac{1}{t-1}+\frac{1}{w-t}\right)
\frac{dw}{dt}  \nn \\
&& + \frac{w(w-1)(w-t)}{8t^2(t-1)^2}\left( \ar-\bb\frac{t}{w^2}
+\gm\frac{t-1}{(w-1)^2}-(\dd-4)\frac{t(t-1)}{(w-t)^2}\right)
\   ,  \eea 
with the identification of the parameters $\ar$,$\bb$,$\gm$,$\dd$ as 
follows:
\bea\label{eq:abcd}
&&\ar=(\mu-\nu+m-n)^2 ~~~~~\         \ ,\      \
\bb=(\mu-\nu-m+n)^2\   ,  \nn \\
&&\gm=(\mu+\nu-m-n-1)^2\     \ ,\      \
\dd=(\mu+\nu+m+n+1)^2\   . 
\eea\ese 
Eq. (\ref{eq:altPVI}) is interesting in its own right since it provides 
us with a covariant way of writing P$_{\rm VI}$, noting its invariance 
under the transformations: 
\[  n\  \ \leftrightarrow\  \ m\     \ ,\      \ 
p\  \ \leftrightarrow\  \ q\     \ ,\      \ 
X\  \leftrightarrow\   1/X\   . \]

\subsection*{N=3:}

This first higher-order case deals with the first genuinely 
multidimensional situation of three two-dimensional sublattices, on 
each of which a copy of the lattice MKdV equation (\ref{eq:genMKdV}) 
is defined. In addition there is also the similarity constraint 
(\ref{eq:pviconstr}) which couples the three lattice directions. 
Thus, for the three-dimensional case we have a coupled system 
of equations whose symbolical representation is shown in figure 2.\\
\begin{figure}[h]
\centering
\epsfig{file=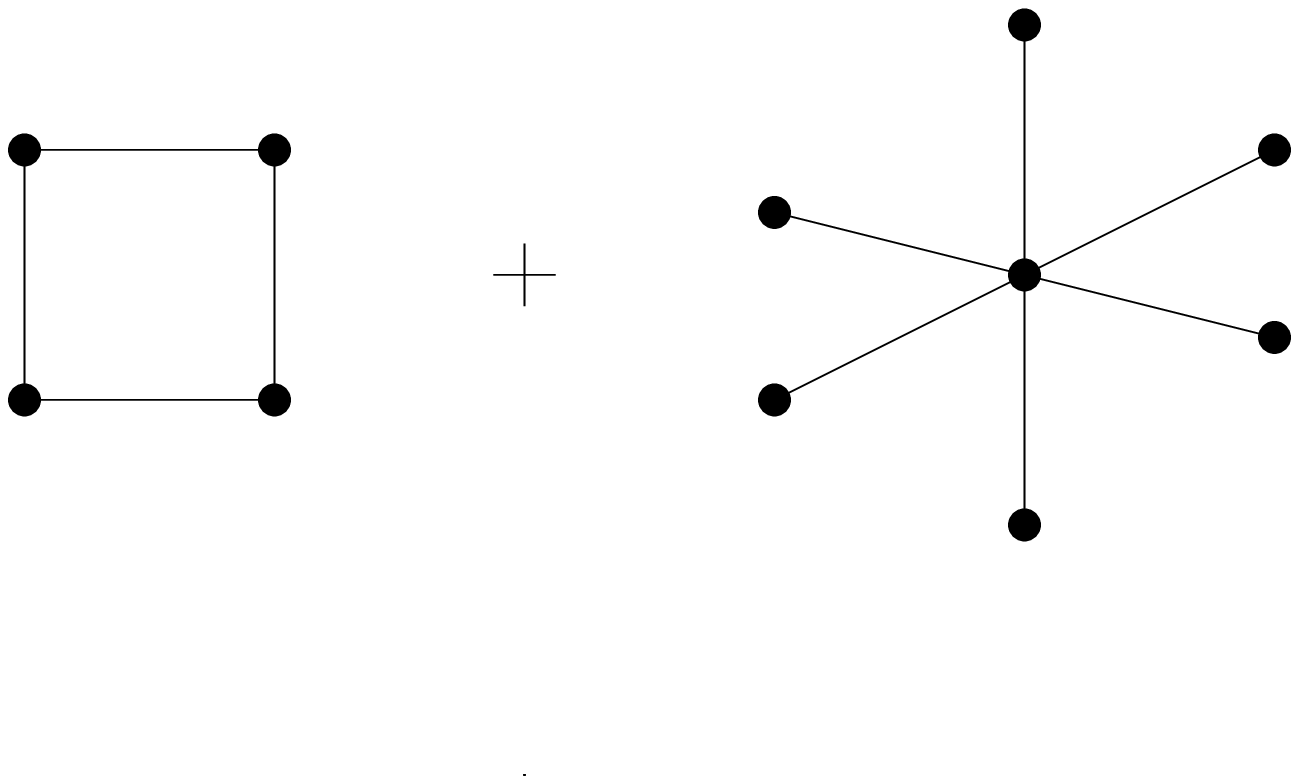,width=3.5in}
\caption{\it{symbolic representation of lattice equation and 
similarity constraint}.}
\end{figure} 
In the previous section we have already demonstrated the consistency 
of the three copies of the lattice equation (\ref{eq:genMKdV}) amongst 
themselves. What remains is to investigate the 
compatability of the lattice equation and the similarity 
constraint, demonstrating that the determination of the values of the 
dependent variable $v$ by using the lattice equation in all three 
directions plus the similarity constraint is unique (assuring the  
single-valuedness of the solution around localised configurations). 

In Figure 3 we have indicated how the iteration of the system proceeds 
starting from a given configuration of initial data (located at 
the vertices indicated by $\bullet$) and moving through 
the lattice by calculating each point by means of either the lattice 
equation (points indicated by $\circ$) or the similarity constraint 
(points indicated by $\times$). The first point where a possible 
conflict arises, due to the fact that the corresponding values of the 
dependent variable can be calculated in more than one way, is 
indicated by $\otimes$. It is at such points 
that the consistency of the similarity reduction 
needs to be verified by explicit computation. This has been carried out 
for this three-dimensional case using MAPLE. Obviously, the iteration 
involves too many steps and the expressions soon become 
too large to reproduce here. 

\begin{figure}[h]
\centering
\epsfig{file=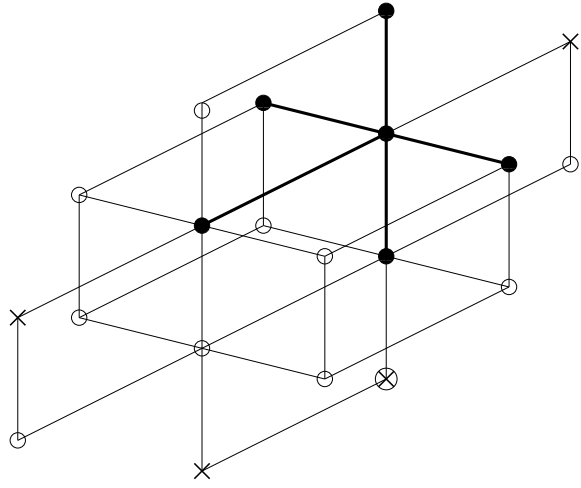,width=3in} 
\caption{\it {Consistency of the constrained lattice system}.}
\end{figure}

In order to analyse the explicit reduction in this case, we 
redefine the following objects
\beast
&a_1=a\ ,\ a_2=b\ ,\ a_3=c&\\
&X_{12}=X\ ,\ X_{13}=Y\ ,\ \mathcal{X}_{12}=\mathcal{X}\ ,\ 
\mathcal{X}_{13}=\mathcal{Y}&
\east
using also $n_1=n$, $n_2=m$, $n_3=h$, as well as 
$p_1=p$, $p_2=q$ and $p_3=r$ to simplify the notation. 
To start with the continuous equations, 
fixing the independent variable to be $p$ we obtain the 
following linear system for the quantities $b$ and $c$ from eq. 
(\ref{eq:plX}) 
\bea
&&\left[ \begin{array}{cc}
               2\mathcal{X} & \mathcal{X}+\mathcal{Y}\\
               \mathcal{X}+\mathcal{Y} & 2\mathcal{Y}
         \end{array} \right]
\left[ \begin{array}{c}
             mb\\
             hc
       \end{array} \right] =
\left[ \begin{array}{c}
             \mu+\nu+p\frac{\pl}{\pl p} \log\,X\\
             \mu+\nu+p\frac{\pl}{\pl p} \log\,Y 
       \end{array} \right] +      \nonumber    \\ 
&&\left[ \begin{array}{cc}
        (\mu-\nu)\mathcal{X}+(n+m)\frac{pq}{p^2-q^2}(\frac{1}{X}-X)
        +h\frac{pr}{p^2-r^2}(\frac{1}{Y}-Y)\\
        (\mu-\nu)\mathcal{Y}+(n+h)\frac{pr}{p^2-r^2}(\frac{1}{Y}-Y)
        +m\frac{pq}{p^2-q^2}(\frac{1}{X}-X)
       \end{array} \right]     \label{eq:bc} 
\eea 
where we have used the similarity constraint to elininate the quantity 
$a$. Furthermore, from (\ref{eq:pla}) we obtain the differential 
relations
\bse\label{eq:dbc}\bea
\frac{\pl (mb)}{\pl p} &=& \frac{mq}{q^2-p^2}[(n+\mu-\nu-mb-hc)
(1-b)\frac{1}{X}\nonumber\\
&&-(1+b)(n-\mu+\nu+mb+hc)X] \label{eq:dbca} \\ 
\frac{\pl (hc)}{\pl p} &=& \frac{hr}{r^2-p^2}[(n+\mu-\nu-mb-hc)
(1-c)\frac{1}{Y}\nonumber\\
&&-(1+c)(n-\mu+\nu+mb+hc)Y]\  .  \label{eq:dbcb}                        
\eea\ese 
Solving $b$ and $c$ from the linear system (\ref{eq:bc}), 
and substituting the results in the differential relations 
(\ref{eq:dbca}) and (\ref{eq:dbcb}), we obtain two coupled 
second-order nonlinear ODE's depending effectively on six free 
parameters, namely $\mu$, $\nu$, $n$, $m$, $h$ and $q/r$. 

Alternatively, we can derive a system of second-order ordinary 
difference equations by fixing one of the discrete variables, 
say $n=n_1$, and using the relations (\ref{eq:arela}) to obtain 
the equations
\bse\label{eq:ddbc}\bea
(pX-q)b+pX+q & = & \frac{(q^2-p^2)X}{qX-p}(\tilde{b}+1) 
\label{eq:ddbca}\\
(pY-r)c+pY+r & = & \frac{(r^2-p^2)Y}{rY-p}(\tilde{c}+1)\   , 
\label{eq:ddbcb} 
\eea\ese 
where the tilde denotes the shift in the lattice direction 
associated with the variable $n$. Using the similarity constraint 
\begin{equation}\label{eq:sconstr} 
na+mb+hc=\mu-\nu\     \ ,\     \ \nu=\ld(-1)^{n+m+h} 
\end{equation}
to eliminate the variables $c$, we 
obtain the following linear system in terms of $\tilde{b}$ and $b$
\bea\label{eq:bb} 
&&\left[ \begin{array}{cc}
       (q^2-p^2)X & -(pX-q)(qX-p)\\
       -m(r^2-p^2)Y & m(pY-r)(rY-p)
       \end{array} \right]
\left[ \begin{array}{c}
       \tilde{b}\\b
       \end{array} \right]=\\ \nonumber
&&\left[ \begin{array}{c}
(pX+q)(qX-p)-(q^2-p^2)X\\
(rY-p)((pY-r)(\mu-\nu-na)+h(pY+r))-(r^2-p^2)Y(h+\mu+\nu-(n+1)\tilde{a})
\end{array} \right]
\eea
where the $a$ and $\tilde{a}$ can be expressed in terms of 
$X$ and $Y$ by
\begin{equation}\label{eq:aa} 
a= \frac{qX\undertilde{X}+p(\undertilde{X}-X)-q}
{qX\undertilde{X}-p(\undertilde{X}+X)+q}
= \frac{rY\undertilde{Y}+p(\undertilde{Y}-Y)-r}
{rY\undertilde{Y}-p(\undertilde{Y}+Y)+q}\  , 
\end{equation}
(where the undertilde denotes the backward shift with respect to 
the discrete variable). 
The system of equations (\ref{eq:ddbc}), (\ref{eq:sconstr}) and 
(\ref{eq:aa})  -- or, equivalently, (\ref{eq:bb}) together with 
(\ref{eq:aa}) leads in principle to a fourth order ordinary 
difference equation in one variable. In fact, solving $b$ and 
$\tilde{b}$ from (\ref{eq:bb}) and then eliminating $b$ altogether 
by a shift in the independent variable $n$ we get a coupled 
system containing one equation in terms of $X$, $\tilde{X}$,
$\tilde{\tilde{X}}$, $\undertilde{X}$ and $Y$, and the equation 
(\ref{eq:aa}) which is first order in the both $X$ and $Y$ with 
respect to the shift in the variable $n$. This system of equations 
depends effectively on six free parameters, namely $\mu$, $\nu$, 
$m$, $h$, $q/p$ and $r/p$.

\section{Isomonodromic Deformation Problem}
\setcounter{equation}{0}
 
The isomonodromic deformation problem for the multidimensional 
lattice system is of Schlesinger type, \cite{Schles}. 
In the two-dimensional case it was already presented in \cite{NP} 
for special values of the parameters $\mu$, $\nu$, 
cf. also \cite{NRGO} for the general parameter case. The extension from 
the two-dimensional to the multidimensional lattice is immediate: one 
only needs to introduce 
additional terms of similar form for each additional lattice direction. 
Thus, the Lax representation consists on the one hand of the linear 
shifts on the lattice of the form 
\begin{equation} 
\psi^i(\kp)=T_i\psi(\kp) = L_i(\kp)\psi(\kp)
\   , \label{eq:Lax} 
\end{equation} 
in which $\kp$ is a spectral parameter, and where the Lax matrices 
$L_i$ are given by 
\begin{equation} \label{eq:MKdVlax}
L_i(\kp)=\left(\begin{array}{cc}
p_i& v^i \\ \frac{\kp}{v}&
p_i\frac{v^i}{v}\end{array}\right) \   , 
\end{equation} 
leading to the Lax equations 
\begin{equation} \label{eq:zerocurv}
L_i^j L_j = L_j^i L_i\   
\end{equation} 
which lead to a copy of the lattice MKdV equation on each 
two-dimensional sublattice labelled by the indices $(i,j)$. On the 
other hand we have the linear differential equation for $\psi(\kp)$ 
with respect to its dependence on the spectral variable $\kp$ 
\bea \label{eq:MKdVmono}
\kp\frac{d}{d\kp}\psi(\kp) &=& \frac{1}{2}\left( \begin{array}{cc}
-(1+\mu)&0\\ 0&\ld(-1)^{\sum_in_i}+\sum_i n_i\end{array} \right)
\psi(\kp) \nn \\ 
&& + \sum_i \frac{n_i\ v}{v^i+\,_{i\!}v} \left(
\begin{array}{cc} 0 & v^i \\  0& -p_i \end{array} \right)
T_i^{-1}\psi(\kp)\   , 
\eea 
the compatibility of which with (\ref{eq:Lax}) leads to the 
similarity constraint (\ref{eq:pviconstr}). 
In addition, we have differential equations for $\psi$ in terms of its 
dependence on the lattice parameters $p_i$ which are of the form 
\begin{equation}\label{eq:MKdVpeq}
\frac{\pl\psi}{\pl p_i} = \frac{n_i}{p_i}\left(
\begin{array}{cc} 1&0\\ 0&0\end{array}\right) \psi
+ \frac{2n_iv}{v^i+\,_{i\!}v} \left( \begin{array}{cc}
0&-\frac{1}{p_i}v^i\\ 0&1\end{array}\right) T_i^{-1}\psi\   ,
\end{equation} 
for each of the variable $p_i$. It is the variables $t_i=p_i^2$ 
that play the role as independent variables in the continuous 
PVI hierarchy. 

The elimination of the back-shifted vectors $T_i^{-1}\psi$ 
by using the inverse of the Lax relations (\ref{eq:Lax}) 
lead to the following linear differential equation for $\psi$ 
\begin{equation}\label{eq:monoexpl}
\frac{\pl\psi}{\pl\kp} = \left(
\frac{A_0}{\kp} + \sum_i \frac{A_i}{\kp-t_i} \right)\psi 
\end{equation} 
thus leading to the problem in the Schlesinger form,  
with regular singularities at $0,\infty,\{t_i\}$. 
The matrices $A_0$ and $A_i$ are given by
\begin{eqnarray*}
&&A_0=\frac{1}{2}\left(\begin{array}{cc}
-(\mu+1) & \sum_i\frac{n_i}{p_i}(1-a_i)v^i\\ 
0 & \ld^{\sum_in_i}+\sum_i n_ia_i\end{array} \right) \  , \\
&& A_i=n_i\left(\begin{array}{cc} \frac{1}{2}(1+a_i) & 
-\frac{1}{2p_i}v^i(1-a_i)\\
-\frac{p_i}{2v^i}(1+a_i)& \frac{1}{2}(1-a_i)\end{array} \right) \   .
\end{eqnarray*}
The continuous isomonodromic deformation is provided by the linear 
differential equations in terms of the lattice parameters, namely
\begin{equation} \label{eq:Peq}
\frac{\pl\psi}{\pl t_i} =
\left( P_i-\frac{A_i}{\kp-t_i}\right)\psi 
\end{equation} 
where
\[  P_i=\frac{n_i}{2p_i}\left(\begin{array}{cc} 
-\frac{1}{p_i}a_i & 0 \\ \frac{1}{v^i}(1+a_i)&0\end{array}\right)\   .\]
Eq. (\ref{eq:Peq}) is not quite in standard form, and we need to 
apply a gauge transformation of the form 
\begin{equation}\label{eq:PVIgauge}
\ol{\psi}\equiv V\psi\    \ ,\    \
V=\left(\begin{array}{cc} 1/v&0\\ U/v&1
\end{array}\right) \   , \end{equation} 
to remove the term with $P_i$, where the auxiliary variable $U$ 
obeys an interesting set of equations by itself (in fact this is the 
object obeying the lattice KdV system of equations) , cf. \cite{NRGO}, 
but we will not give any details here. With this gauge, the continuous 
isomonodromic deformation (\ref{eq:Peq}) adopts the standard form
\begin{equation}\label{eq:PVIstandard}
\frac{\pl\ol{\psi}}{\pl t_i}=-\frac{\ol{A}_i}{\kappa-t_i}
\ol{\psi}
\      \ , \     \  \ol{A}_i=V A_i V^{-1}\   ,
\end{equation} 
whilst the discrete isomonodromic is readily obtained from the Lax 
representation of (\ref{eq:Lax}). 

\section{Connection with Garnier Systems} 
\setcounter{equation}{0}

Interestingly, already M.R. Garnier in his seminal paper of 1912, 
\cite{Garnier}, embarked on the question of finding higher-order 
analogues of the PVI equation, adopting the method that was proposed 
somewhat earlier by R. Fuchs, in \cite{Fuchs}, which can be identified 
with the isomonodromic deformation approach, cf. also \cite{Schles}. 
Garnier gave a general construction of such higher-order equations 
constituting coupled systems of partial differential equations, which 
are the isomonodromic {\it Garnier systems}. As a particular 
example, he wrote down explicitely in \cite{Garnier} the first 
higher-order PVI equation in terms of the following coupled system,  
consisting of the second order ODE in terms of two dependent variables 
$w=w(t,s)$ and $z=z(t,s)$ 
\bse \label{eq:hiPVI}\bea \label{eq:hiPVIa} 
\frac{\pl^2 w}{\pl t^2} &=& \frac{1}{2}\left( \frac{1}{w}+
\frac{1}{w-1}+ \frac{1}{w-t}+ \frac{1}{w-s} - \frac{1}{w-z}\right)
\left(\frac{\pl w}{\pl t}\right)^2 \nn \\ 
&& - \left(\frac{1}{t}+ \frac{1}{t-1}+ \frac{1}{t-s}- \frac{1}{t-w}-
\frac{1}{t-z}\right) \frac{\pl w}{\pl t}  \nn \\
&& + \frac{1}{2}\frac{w(w-1)(w-s)(z-t)}{z(z-1)(z-s)(w-t)(z-w)}
\left(\frac{\pl z}{\pl t}\right)^2 - \frac{w-t}{(z-t)(z-w)} 
\left(\frac{\pl w}{\pl t}\right)\left(\frac{\pl z}{\pl t}\right)
\nn \\ 
&& +\frac{2w(w-1)(w-t)(w-s)(z-t)^2}{t^2(t-1)^2(t-s)^2(w-z)}\times\nn\\  
&& ~~~~~~~ \times \left[ \ar+\bb+\gm+\dd+\kp +\frac{7}{4} 
-\frac{ts}{z}\,\frac{\ar+\frac{1}{4}}{w^2}+\frac{(t-1)(s-1)}{(z-1)}\, 
\frac{\bb+\frac{1}{4}}{(w-1)^2} \right. \nn \\ 
&& ~~~~~~~~~~~~ \left. + \frac{t(t-1)(t-s)}{(z-t)}\, 
\frac{\gm}{(w-t)^2} + \frac{s(s-1)(s-t)}{(z-s)}\, 
\frac{\dd}{(w-s)^2} \right] \nn \\ 
\eea
together with coupled first order PDE's 
\bea\label{eq:hiPVIb} 
\frac{t(t-1)}{t-z}\frac{\pl w}{\pl t}
+\frac{s(s-1)}{s-z}\frac{\pl w}{\pl s} = \frac{w(w-1)}{w-z}\  ,   \\ 
\frac{t(t-1)}{t-w}\frac{\pl z}{\pl t}
+\frac{s(s-1)}{s-w}\frac{\pl z}{\pl s} = \frac{z(z-1)}{z-w}\  .   
\label{eq:hiPVIc}\eea \ese 
It should be pointed out that the system consisting of 
(\ref{eq:hiPVIa}), (\ref{eq:hiPVIb}) and (\ref{eq:hiPVIc}) amounts 
actually to a fourth order ODE in terms of $w=w(t)$ only, and as such 
can be rightly considered to be the first higher-order member of 
the Painlev\'e VI hierarchy. In fact, Garnier gave in his paper a 
number of important assertions: {\it i)} that his system of equations 
is  completely integrable\footnote{Obviously, Garnier's use of the 
term  {\it integrability}  was meant here in the precise sense of 
that of a compatible system, very much in the same sense as the 
compatibility of the continuous and discrete systems that we have 
encountered in sections 2 and 3.}, and that it degenerates (under the
autonomous limit) to a system living 
on the Jacobian of a hypereliptic curve, {\it ii)} that the symmetric 
combinations of the {\it dependent} variables of the system, as 
functions of {\it each one} of the essential singularities (i.e. 
singling out one of the independent variables) are meromorphic in terms 
this variable except for the fixed critical points which are at 0,1,
$\infty$, or at the location of the values of the other independent 
variables\footnote{This assertion amounts to the well-known 
Painlev\'e property.}, {\it iii)} that for the parameters of the system 
in general position 
the symmetric functions of the dependent variables are essentially 
transcendental functions of the constants of integration (i.e. of the 
initial data).  

Subsequent work on the Garnier systems was done mostly by K. Okamoto 
and his school, cf. e.g. \cite{Oka,IKSY}. However, it seems that 
in most of these works these systems were treated rather as an 
underdetermined system of PDE's rather than (as Garnier himself  
clearly had in mind) as a consist system of ODE's. Although it is not 
easy to find the explicit transformation of the lattice system 
exposed in sections 2 and 3 to the systems that Garnier wrote down, 
in particular to find the explicit relation between the above system 
(\ref{eq:hiPVI}) and the system consisting of (\ref{eq:bc}) and 
(\ref{eq:dbc}), it is to be expected that such a mapping exist. The 
identification is probably easiest to obtain via the transformation of 
the corresponding 
Schlesinger type of system as given in section 4 and the linear 
system that Garnier exploited in \cite{Garnier}. However, the search 
for such an identification will be left to a future study. 

Let us finish with some remarks on the relevance of these results 
for work that is been done in recent years. One of the most 
exciting developments is the way in which the issue of algebraic 
solutions of P$_{\rm VI}$ have arisen in recent years, e.g. 
in connection with WDVV equations, Frobenius manifolds and 
quantum cohomology, cf. e.g. the review \cite{Dubr2}. Such algebraic 
solutions were already known to Picard, Painlev\'e and Chazy. 
In fact, in his early paper \cite{Fuchs} R. Fuchs obtained 
a realisation of P$_{\rm VI}$ in terms of an elliptic integral, and 
this realisation was subsequently used by Painlev\'e in \cite{Pain}, 
to derive an elliptic form for the P$_{\rm VI}$ equation, a form of 
the equation that was recently recovered by Manin in \cite{Manin}
\footnote{We are grateful to R. Conte for pointing out the reference 
\cite{Pain}.}. The assertions of Garnier in \cite{Garnier} on his 
generalisation of the Fuchs' approach might form a starting point for 
extending this elliptic connection to the Garnier systems, in which 
case we would expect to be able to find a realisation of those systems 
in terms of hyperelliptic integrals rather than elliptic ones. 
This might eventually lead to the construction of algebraic 
solutions of those systems, possibly in the spirit of the recent 
papers \cite{Korotkin,Deift}. It would be of interest to further 
investigate the role of the discrete systems in connection with 
the Garnier systems: we expect them to constitute the superposition 
formulae for the underlying higher root systems of the 
corresponding affine Weyl groups. Thus, eventually, a geometric 
interpretation of the Garnier systems and their discrete analogues 
in the sense of the blowings-up of the corresponding rational 
surfaces of their initial conditions, along the lines of the recent 
paper \cite{Sakai}, might be anticipated.


\begin{thebibliography}{99}
\bibitem{Carg}
B. Grammaticos, F.W. Nijhoff and A. Ramani, Discrete Painlev\'e
Equations, {\it The Painlev\'e
Property, One Century Later}, Edited by R. Conte, CRM series in 
mathematical physics, Springer Verlag, 1999, pp. 413-516. 
\bibitem{Sakai} 
H. Sakai,  Rational surfaces associated with affine root systems 
and geometry of the Painlev\'e equations, Preprint Kyoto university, 
May 1999. 
\bibitem{NRGO}
F. W. Nijhoff, A. Ramani, B. Grammaticos and Y. Ohta, On Discrete 
Painlev\'e Equations Associated with the Lattice KdV Systems and 
the Painlev\'e VI Equation, {\tt solv-int/9812011}, (submitted to 
Stud. Appl. Math.). 
\bibitem{NP}
F.W. Nijhoff and V.G. Papageorgiou, Similarity Reductions of Integrable
Lattices and Discrete Analogues of the Painlev\'e II Equation,
{\it Phys. Lett.} A153:337--344 (1991).
\bibitem{DIGP}
F.W. Nijhoff, Discrete Painlev\'e Equations and Symmetry Reduction
on the Lattice, in: {\it Discrete Integrable Geometry and Physics}, 
eds.  A.I. Bobenko and R. Seiler, (Oxford Univ. Press, in press). 
\bibitem{NJH} 
F.W. Nijhoff, N. Joshi and A. Hone, On the discrete and continuous 
Miura chain associated with the Sixth Painlev\'e equation, 
{\tt solv-int/9906006}, {\it Phys. Lett.} {\bf A} in press. 
\bibitem{NHJ}
F.W. Nijhoff, A. Hone and N. Joshi, On a Schwarzian PDE Associated 
with the KdV Hierarchy, {\tt solv-int/9909026}.
\bibitem{FN}
H. Flachka and A.C. Newell, Monodromy- and spectrum-preserving 
deformations I, {\it Commun. Math. Phys.} {\bf 76}:65--116 (1980). 
\bibitem{Kudry}
N. A. Kudryashov, The first and second Painlev\'e equations of higher 
order and some relations between them, {\it Phys. Lett.} A224:353--360 
(1997).   
\bibitem{Cres}
C. Creswell and N. Joshi, The Discrete Painlev\'e I Hierarchy, in: 
{\it Symmetries and Integrability of Difference Equations}, eds. 
P. A. Clarkson and F. W. Nijhoff, (Cambridge univ. Press, 1999), pp. 
197--205. 
\bibitem{Hone}
A. N. W. Hone, Non-autonomous H\'enon-Heiles systems, {\it Physica} 
D118:1--16 (1998). 
\bibitem{Noumi}
M. Noumi and Y. Yamada, Higher Order Painlev\'e Equations of Type 
$A_l^{(1)}$, {\tt math.QA/9808003}. 
\bibitem{Pick}
P. A. Clarkson, N. Joshi and A. Pickering, B\"acklund transformations 
for the second Painlev\'e hierarchy: a modified truncation approach, 
{\tt solv-int/9811014}. 
\bibitem{Kitaev}
A. V. Kitaev, Elliptic Asymptotics of the first and the second 
Painlev\'e transcendents, {Russ. Math, Surv.} 49:81--150 (1994). 
\bibitem{Garnier}
M.R. Garnier, Sur des \'equations diff\'erentielles du troisi\`eme 
ordre dont l'int\'egrale g\'en\'erale est uniforme et sur une 
classe d'\'equations nouvelles d'ordre sup\'erieur, {\it Ann. 
\'Ecol. Norm. Sup.}, vol. 29:1--126 (1912). 
\bibitem{Fuchs}
R. Fuchs, Sur quelques \'equations differentielles lineaires du second 
ordre, {\em C. R. Acad. Sci. (Paris)} 141:555--558 (1905).
{\it Math. Ann.} 63:301--321 (1907).
\bibitem{Hitch}
N. J. Hitchin, Poncelet Polygons and the Painlev\'e Transcendents, 
in: {\it Geometry and Analysis}, ed. Ramanan, (Oxford Univ. Press, 
1995), pp. 151--185.
\bibitem{Manin} 
Yu. I. Manin, Sixth Painlev\'e Equation, Universal Elliptic Curve 
and Mirror of $\mathbb P^2$, {\tt alg-geom/9605010}. 
\bibitem{Olsh}
A. M. Levin and M. A. Olshanetsky, Painlev\'e--Calogero 
correspondence,  Preprint ITEP-TH23/97, {\tt alg-geom/9706010}.  
\bibitem{Dubr}
B. Dubrovin and M. Mazzocco, Monodromy of certain Painlev\'e-VI 
transcendents and reflecion groups, {\tt math.AG/9806056}. 
\bibitem{KDV}
F.W. Nijhoff and H.W. Capel, The Discrete Korteweg-de Vries Equation,
{\it Acta Applicandae Mathematicae} 39:133--158 (1995).
\bibitem{Schles} 
L. Schlesinger, \"Uber eine Klasse von Differentialsystemen beliebiger 
Ordnung mit festen kritischen Punkten, {\it J. f\"ur Math.} {\bf 141}:
96--145 (1912). 
\bibitem{Oka}
K. Okamoto, Isomonodromic deformation and Painlev\'e equations, and the 
Garnier System, {\it J. Fac. Sci. Univ. Tokyo} IA, 33 (1986), 575-618.
\bibitem{IKSY}
K. Iwasaki, H. Kimura, S. Shimomura and M. Yoshida, {\em From
Gauss to Painlev\'e. A Modern theory of Special Functions},
Vieweg Verlag, Braunschweig, 1991, Ch. 3.
\bibitem{Dubr2}
B. Dubrovin, Painlev\'e transcendents in two-dimensional topological 
field theory, in: {\it The Painlev\'e
Property, One Century Later}, Edited by R. Conte, CRM series in 
mathematical physics, Springer Verlag, 1999, pp. 287-412. 
\bibitem{Pain}
P. Painlev\'e, Sur les \'equations diff\'erentielles du second ordre 
\`a points critiques fixes, {\it C.R. Acad. Sci.} (Paris) {\bf 143} 
\#26 (1906) 1111-1117.
\bibitem{Korotkin}
A. V. Kitaev and D. A. Korotkin, On solutions of the Schlesinger 
Equations in terms of $\Theta$-functions, {\it math-ph/9810007}. 
\bibitem{Deift}
P. Deift, A. Its, A. Kapaev and X. Zhou, On the Algebra-Geometric 
Integration of the Schlesinger Equations, Preprint Courant Institute, 
1999. 
\end{thebibliography}
\end{document}